# Synthesis of Arbitrary Quantum Circuits to Topological Assembly: Systematic, Online and Compact

Alexandru Paler[1], Austin G. Fowler[2], and Robert Wille[3]

[1]Johannes Kepler University / Linz Institute of Technology, Linz 4040, Austria
[2]Google Inc., Santa Barbara, California 93117, USA
[3]Johannes Kepler University, Linz 4040, Austria
*alexandru.paler@jku.at

**ABSTRACT**

It is challenging to transform an arbitrary quantum circuit into a form protected by surface code quantum error correcting codes (a variant of topological quantum error correction), especially if the goal is to minimise overhead. One of the issues is the efficient placement of magic state distillation sub circuits, so-called distillation boxes, in the space-time volume that abstracts the computation's required resources. This work presents a general, systematic, online method for the synthesis of such circuits. Distillation box placement is controlled by so-called schedulers. The work introduces a greedy scheduler generating compact box placements. The implemented software, whose source code is available at www.github.com/alexandrupaler/tqec, is used to illustrate and discuss synthesis examples. Synthesis and optimisation improvements are proposed.

## Introduction

Synthesis and optimisation of topologically error corrected quantum circuits is ongoing research. The first complete framework for the synthesis of arbitrary quantum circuits to a topological assembly was presented in[1]. Synthesis efficiency was not addressed at that time, and no attempts were made until now. The circuits resulting from[1] included a large number of fault-tolerant qubit initialisations inserted before the execution of any quantum gate. This implied that large portions of the quantum computer's hardware was used only briefly, and remained unused for almost the entire duration of the circuit's execution. The following sections offer a condensed gradual introduction to topologically error corrected quantum circuits. This is required in order to motivate the required improvement of the state-of-the-art constituted by[1].

## Background

Arbitrary quantum computations are generally formulated as a quantum circuit consisting of quantum operations (initialisations, quantum gates, measurements) and quantum wires (e.g. Fig. 1). A time axis can be associated to the circuit's execution, and time flows from left to right. A circuit is executed sequentially with regard to the ordering of operations on each quantum wire, so that preceding quantum operations are always on the left side of the currently executed operation. Succeeding operations are on the right side. Quantum circuit wires cannot run backwards in time, because, although quantum circuits have a two dimensional representation, the wires connecting the quantum operations have a temporal and not a hardware interpretation (classic circuit wires are mostly associated to hardware).

Circuit execution depends on the availability of computational resources, which for a quantum computer are time and hardware: the time needed to operate the hardware in order to execute the entire computation. It can be assumed that hardware is arranged in a two dimensional lattice, so that the overall amount of available resources is abstracted by a metric similar to a *space-time volume* resulting after multiplying available hardware (strictly limited) with available time (faster is better).

Throughout this work (e.g. Fig. 1), the three axes in each figure indicate the flow of time (green) and the two dimensional arrangement of the hardware expected to be used (red and blue)[6]. The temporal axis associated to a circuit's execution is indicated by the green axis.

*Motivation*
The goal of this work is to take an arbitrary Clifford+T quantum circuit (maximum 100 qubits) and to *synthesise* (convert) it into a fault-tolerant form performing the same computation protected by a specific variant of topological quantum error correction code (QECC), the surface QECC[4,5]. The synthesis result, called *topological assembly*, should be *compact* with regard to the required computational resources.

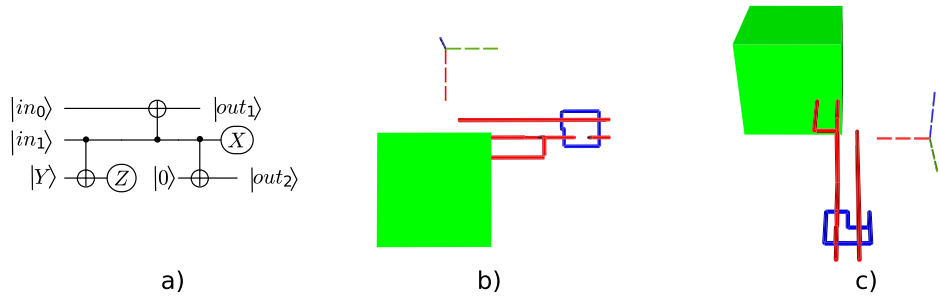

**Figure 1.** Different perspectives of the quantum circuit and the resulting geometric description. The middle and right figures illustrate a distillation box (green), defects (red and blue) and three braids. In a) and b) circuit execution starts from left and proceeds to the right, and in c) the execution starts from the back towards the front of the figure.

Prior to circuit execution, synthesis transforms the Clifford+T circuit into a fault-tolerant form which gets afterwards protected by the surface QECC. Circuit execution requires on the fly (*online*) synthesis methods, which in this work follow a certain pattern, they are *systematic*.

*Fault-Tolerant Quantum Circuits*
There are three aspects of fault-tolerant quantum computations[2]. First, the number of errors suffered by the hardware should be limited below a QECC specific threshold, allowing scalable computations to be achieved by dedicating more hardware to the QECC. Second, no matter where an error occurs, it must not be multiplied by subsequent gates such that its effect on the computation becomes uncontrolled. Third, certain sub-computations need to be repeated until an acceptance signal is generated, and there must be a mechanism to detect this signal and schedule appropriate repetitions as the quantum computation proceeds, i.e. in an online manner. While much experimental quantum computing work has been done, there has to date been no demonstration of even the first requirement for fault tolerant quantum computation, namely a universal set of gates with sufficiently low error rates to even in principle successfully scale up any given technology. From a practical perspective, however, even such a demonstration may not be sufficient, as the total cost of hardware needed to achieve a practical logical error rate throughout an algorithm of interest could be a significant real-world barrier to the construction of a quantum computer. There are two general philosophies one could adopt when trying to minimise the hardware cost, either build many low fidelity but cheap quantum systems, or fewer high fidelity but more expensive systems. At this point in time, it is not clear even which category most experimental efforts are likely to fit into.

The synthesis input circuit uses Clifford+T gates, which are decomposed into an ICM form[3] (the circuit consists entirely of single qubit initialisations, CNOT and single qubit measurements). In practice, the ICM form is achieved by implementing gates through teleportation sub circuits. Such sub circuits are probabilistic and introduce an online/dynamic behaviour in the overall circuit structure.

Teleported gates are, in this work, expressed using only the states $|0\rangle$, $|+\rangle$, $|A\rangle = \frac{1}{\sqrt{2}}(|0\rangle + e^{i\frac{\pi}{4}}|1\rangle)$, $|Y\rangle = \frac{1}{\sqrt{2}}(|0\rangle + i|1\rangle)$, together with CNOT gates and single qubit measurements (*X* and *Z* basis). The second step is to optimise the ICM circuit, e.g. using wire recycling[20].

## Topological Quantum Error Correction
The optimised ICM form of the input circuit is protected, at this stage, with the surface QECC.

*Defect*
A defect is the representation of a sequence of low level quantum hardware operations required for protecting the state of a circuit qubit. By definition, a defect can be either primal or dual. A logical qubit is formed by pairs of same type defects and, as a result, the surface QECC allows the construction of primal and dual logical qubits.

*Braid*
A logical CNOT gate is obtained by braiding between defects of opposite type (a primal and a dual): the dual logical qubit controls the primal logical qubit (target). Braids between defects of the same type leave the corresponding logical qubit states untransformed; the result is a logical identity gate.

*Distillation box*
The surface QECC cannot ensure that an arbitrary initialised qubit will have a high fidelity. As a result, fidelity is increased by distillation procedures expressed as sub circuits. Such procedures take multiple low fidelity instances of a state and output a single high fidelity state. Consequently, a code will protect circuits including distillation procedures (sub circuits). The resulting circuit includes distillations, symbolised by boxes of different types ($|A\rangle$ or $|Y\rangle$), where each box is a placeholder of the protected distillation sub circuit. Distillations are probabilistic (may not succeed and the output state has low fidelity) and heralded (it is known if distillation succeeded).

*Geometric description*

The result of protecting the ICM circuit with the surface code is a *geometric description* (of the topological assembly), and it consists *entirely* of single or pairs of point-like coordinates from the space-time volume. Coordinate are used for representing: circuit inputs and outputs, each straight defect segment describing CNOTs and identity gates, the position where distilled magic states need to be delivered to in the circuit, and where distillation boxes need to placed in the space-time volume. Except the latter, all coordinates can be computed offline, before circuit execution. Circuit execution requires the online computation of box placement and how these are connected to the circuit through defects.

For example, all the elements from Fig. 2 can be represented using point-like coordinates. A pair of coordinates determines two diagonal corners of each yellow or green box. Fig. 3 shows that all yellow, blue or red defects consist of multiple linear segments: a pair of coordinates is sufficient for the two end points of each segment.

The figures presented in this work are geometric description visualisations (e.g. after boxes were placed, segments were drawn) obtained from the space-time coordinates resulting after protecting the ICM circuit with surface codes. In the following, all the introduced concepts (e.g. connection, connection pool) have a geometric representation, because their implementation in the circuit consist of defects, which are sequences of linear segments having end points represented by point-like space-time volume coordinates.

## Execution of Online Synthesised Circuits

The previously generated geometric description needs to be translated into specific quantum gates for the quantum hardware. However, the probabilistic nature of fault-tolerant operations and the amount of available computational resources play a role during circuit execution. This influences the placement of the distillation boxes and how they are connected to the circuit.

Therefore, during online synthesis, distillation boxes are *scheduled* (placed) into the space-time volume associated to the quantum computer. A scheduling round, is the time period when boxes are scheduled, and a *distillation layer* is the space-time region where distillation boxes were placed during a round. Various strategies exist to ensure that sufficient boxes succeed in a round: either boxes are repeatedly executed (placed along the time axis), or, if sufficient computational resources are available, multiple boxes are executed in parallel and only successful ones are used.

A *schedule* specifies the point-like coordinates of each distillation box from a distillation layer. A synthesised circuit is generated after multiple rounds and, includes multiple distillation layers with their corresponding schedules.

Circuit execution is continued after each scheduling round. Synthesis uses feedback from the quantum hardware to determine whether specific boxes have been successful, and the geometric description is then extended with appropriate *connections* (pairs of defects) from the successful distillation boxes to the places where distilled states are required. The connections in turn are converted to specific gates for the quantum hardware.

A systematic construction of connections is achieved by introducing a *connection pool*, where the outputs of successful distillations are stored/placed until these are required by the circuit. The pool intermediates the process of constructing connections. Its functionality is described in the *Methods* section.

## Technical Motivation

There are many types of topological quantum error correction, and the herein detailed synthesis is not a formalism that encompasses all of them. The foundation of topological QECCs were formulated in[7,8,24], and the braiding (also called code deformation) methods were introduced in[9,10]. The application of distillation procedures was initially proposed in[11], and supported the definition of universal quantum computations protected by topological QECCs.

Distillation procedures are considered a source of high computational resource consumption. As a result, multiple proposals were made in order to reduce the amount of necessary distillations. The approaches range from recycling $|Y\rangle$ distillations in the surface code[5], to using other kinds of topological codes (e.g. colour variants)[12,13] or implementing topological twists (e.g.[14]) in the surface code in order to totally eliminate the need for $|Y\rangle$ distillations.

This work, as well as[1], takes a conservative approach dictated by the design characteristics of the first generation of envisioned quantum hardware: nearest neighbour interactions between physical qubits are allowed in a two dimensional lattice supporting quantum gates and quantum measurements sufficient to implement only the classic surface QECC. Furthermore, no studies were undertaken to investigate the resource efficiency of reusing or removing $|Y\rangle$ distillations. For example, it is not obvious that recycling the former, as shown in[5], would lead to lower overhead as then a pair of CNOTs and a pair of Hadamard gates plus routing of the recycled states is required. On the contrary, the compact version of a $|Y\rangle$ distillation has an overhead potentially as low as 1.5 the minimum volume CNOTs[15] plus routing. Additionally, no explicit Hadamard has been devised for when using surface QECCs with topological cluster states. This conservative approach is motivated by practical perspectives, too. The practicality of non-surface topological codes has not been studied to show how to overcome their non local nature in two dimensions, low thresholds, and identical complexity overhead scaling to state distillation. As a result, distillation procedures can be theoretically circumvented, but there is ongoing research on how to apply this into practice.

The various flavours of synthesis need also to be discussed, in order to motivate this work. Synthesis of quantum gates is to express an arbitrary unitary operation into a sequence of gates from a discrete set (usually Clifford+T) that approximates within a reasonable error the initial unitary. One of the major costs to optimise synthesis results is to minimise the *T-count*, the number of T gates from the decomposition. This is important with respect to surface code QECCs, because each T gate necessitates $|A\rangle$ state distillations. Efforts at reducing T-count are well documented and range from heuristic and general approaches (e.g.[16,17]) to ones targeting particular gates (e.g.[18]).

There are multiple synthesis steps involved in preparing an arbitrary quantum circuit for execution, and this works considers circuits being already decomposed into Clifford+T sequences, and furthermore into a fault-tolerant form called ICM (see previous section). From this perspective, the current synthesis shares some similarities to online place and route methods[19]: circuit elements need to be placed into the space-time volume, and connections between the elements need to be dynamically computed (routed).

## Results

This work advances the state-of-the-art of synthesis by constructing a systematic algorithm to place compact layers of distillation boxes around standardised braids driving the circuit logic. Distillations were modelled with a failure probability, and the necessary pathways connecting successful distillations to where they are needed were constructed only after this information was obtained. The new framework does not take into account the approach of the end of the circuit, where fewer distillations would be required, however in practice this would not affect the overhead of a large circuit.

The synthesis framework built according to the following requirements: 1) there is no general restriction on the placement of distillation boxes (in contrast to[1]), as long they do not overlap with other circuit elements; 2) distillation box connections should be computed using a path finding algorithm instead of having a fixed structure like in[1].

Throughout this work, distillation failure probability was set to 50%. This parameter value is arbitrary, and was chosen to highlight the systematic and online functionality of the synthesis. Choosing a lower failure probability (which is more realistic) will result in fewer boxes being scheduled (lower space-time overhead introduced by distillations) and a significantly higher ratio of connections to be computed. Thus, connection defects are more dense in the space-time volume occupied by the circuit, and this impacts the illustrations from this work as well as the existing simulation results: the improvement against the previous state of the art synthesis would be much higher, not only because the current work delivers more compact space-time volumes, but because the previous one was very naive.

Distillation failure probability influences the probability of successfully executing a protected circuit (reliability), which in turn impacts the resource overhead. The pursuit of even more aggressive optimisation (see Discussion section) is sensitive to distillation failure probability, but, more important, to the overall reliability required for a computation: high reliability necessitates even more efficiency improvements in the synthesis and optimisation algorithms.

The ICM Toffoli circuit from Fig. 2 is an example of what the improved synthesis generates. The entire ICM circuit implements the functionality of a single Toffoli gate. The choice of the circuit is not entirely arbitrary, because multiple quantum arithmetic circuits are based only on Toffoli gates. Investigating the resources required to implement a single Toffoli gate is an indication of the difficulty to implement fault-tolerant quantum arithmetic computations.

Fig. 2 both types of distillation boxes are represented (yellow for $|Y\rangle$ and green for $|A\rangle$), and there are five vertical distillation layers accompanying the circuit. The time axis indicating circuit execution runs from the left to the right. The geometric description of the ICM circuit for which the topological assembly was generated is visible, for example, between scheduling layers two and three. The geometric description is the lowest horizontal layer of blue defects.

Compared to previous state-of-the-art methods, a more compact geometric description is obtained by placing distillation boxes as close as possible to the latest time point when a qubit initialisation is needed: a successful distillation output should be available each time an initialisation is required. Although the new framework supports flexible box placement, the current synthesis achieves compactness by using a specific box scheduler which introduces a kind of placement restriction.

Fig. 2 is complex, but it can be decomposed into constituent elements. The following figures use a different colour coding. Fig. 3 shows the connections (yellow) between boxes and connection pool (blue), and the geometric description (red). The scheduling rounds and the boxes (irrespective of the their type) are drawn in Fig. 4. A simplified illustration of the elements is provided in Fig. 5. Each element of the synthesis result is handled by an individual framework component. The interactions between the components are presented in the following.

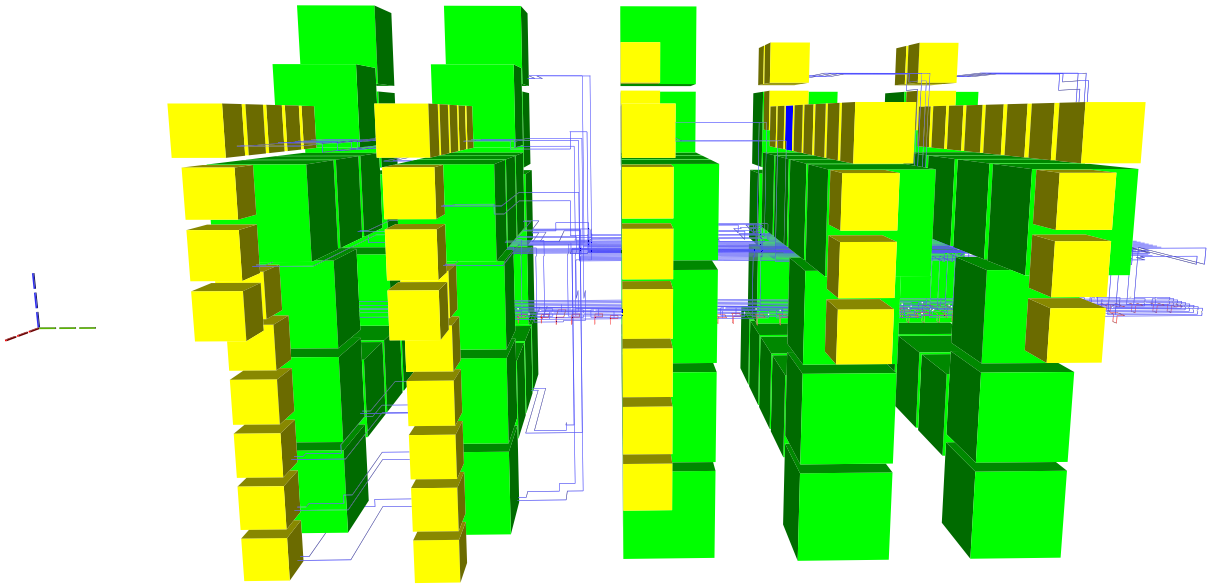

**Figure 2.** The synthesis result of the optimised ICM implementation of the Toffoli gate.

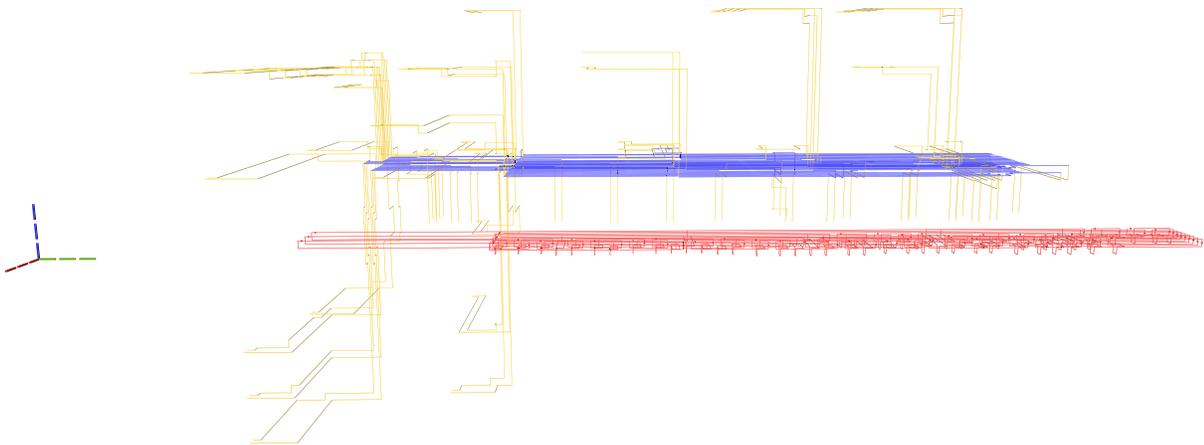

**Figure 3.** A synthesis result consists of defects and boxes. For example, the defects from Fig. 2 include the connections (yellow), connection pool (blue) and geometric description (red).

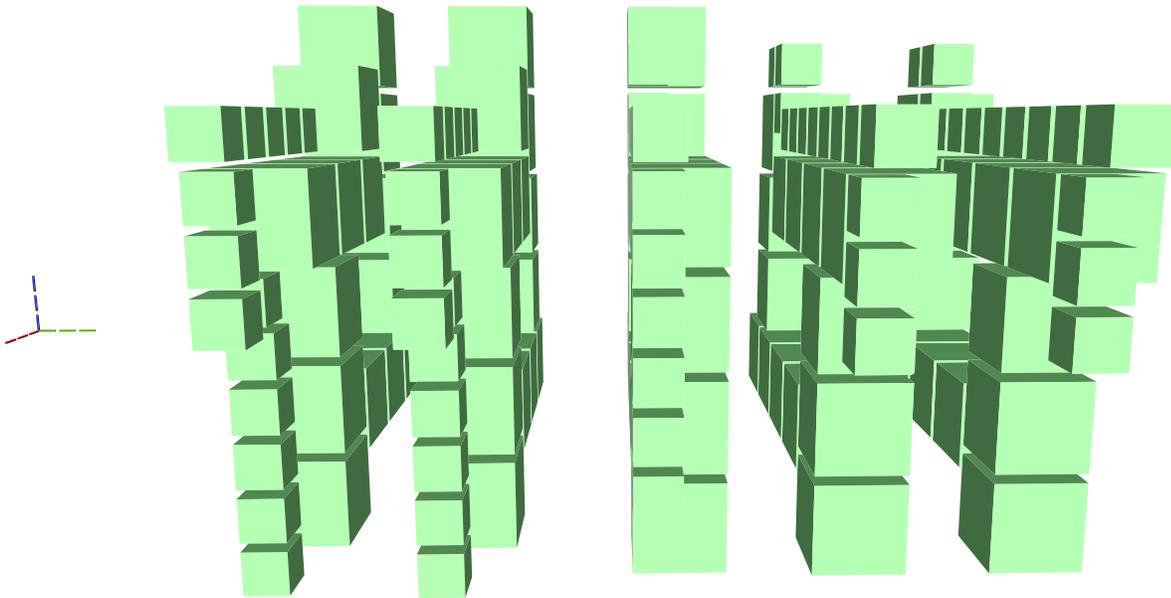

**Figure 4.** A synthesis result consists of defects and boxes. The boxes (green, box type differences are not highlighted) from Fig. 2 are arranged into scheduling rounds (five in this example). Each round is executed at a specific point in time (green dotted axis), and separate rounds are separated in time. The yellow defects from Fig. 3 are constructed only between rounds.

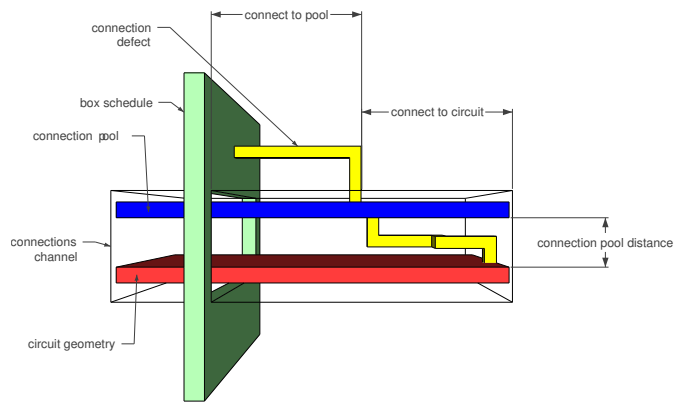

**Figure 5.** Simplified view of the elements forming a scheduling result. Some of the parameters controlling the synthesis are indicated.

**6/18**

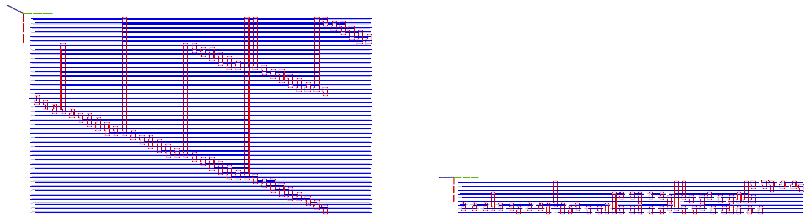

**Figure 6.** The geometric description of the ICM Toffoli gate: left) the unoptimised version; right) the optimised version.

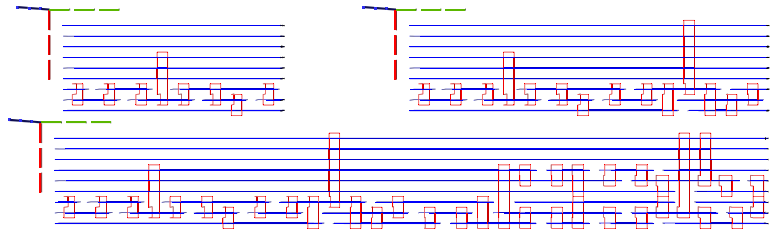

**Figure 7.** Three steps of the optimised circuit (Fig. 6) traversal. The traversal (growth) of the circuit is controlled by the inputs requiring high fidelity initialisation states. Circuit traversal is interrupted each time at least such an input is encountered. The geometric description is updated to the latest time point before the inputs: the right most side of each geometric description does not include them. The encountered inputs are triggering scheduling events, and are used for the geometric description only in the next step. Therefore, the inputs that stopped the step from the left figure are existing in the following geometric descriptions (e.g. middle and right figures).

### Geometric description generator
The entire QECC protected circuit can be geometrically described. Braided defect pairs are continuously generated according to the ICM circuit form. Scheduling events could influence the layout of the ICM form geometric description, if scheduling is allowed to delay circuit initialisations.

### Quantum circuit optimiser
Synthesis starts by loading the quantum circuit required to be synthesised into a topological assembly and, if required, transform it into ICM. One of the most effective ICM quantum circuit optimisation strategies has proven to be, for the moment, wire recycling[20]. Due to their nature, such circuits, include a very high number of wires (qubits). However, each wire is operated on by only a few CNOT gates. Furthermore, in addition to CNOT gates, ICM uses only specific single-qubit initialisations and measurements. Quantum circuit wires and qubits are distinct concepts, and it is possible to reuse the same wire for initialising and measuring a series of qubits. Many of the logical qubits in the left figure are initialised later and measured earlier. After this is done, the same horizontal line can support the sequential initialisation and measurement of multiple logical qubits. It was possible to reduce the number of wires by up to 90% while maintaining the computation unchanged (number of qubits and ordering of initialisations, CNOTs and measurements)[21]. An optimisation example is presented in Fig. 6.

### Event driven circuit traversal
Synthesis and execution are an online, event based traversal of the corresponding ICM quantum circuit. It is straightforward to generate the geometric description of any ICM operation except for $|A\rangle$ and $|Y\rangle$ initialisations, because these depend on successful distillation boxes. Circuit structural dynamism results from the need to achieve fault-tolerant state initialisations. The synthesis of the geometric description cannot be performed in a single step, because each qubit initialisation requires an appropriate high fidelity state. Therefore, the ICM circuit is iteratively traversed in a sequence of steps, each triggering a scheduling event whenever an initialisation is encountered (e.g. Fig.7). A qubit can be initialised only if successful distillation box outputs are available.

### Distillation box scheduling
Scheduling events are treated by placing distillation boxes into the space-time volume of the topological assembly. The result of each event is a distillation layer with an accompanying schedule.

All ICM operations preceding the initialisation are executed. For example, in Fig. 2, the first distillation layer is scheduled after a traversal step of the corresponding ICM circuit encountered $|A\rangle$ and $|Y\rangle$ initialisations. The geometric description is generated up to those input time coordinates.

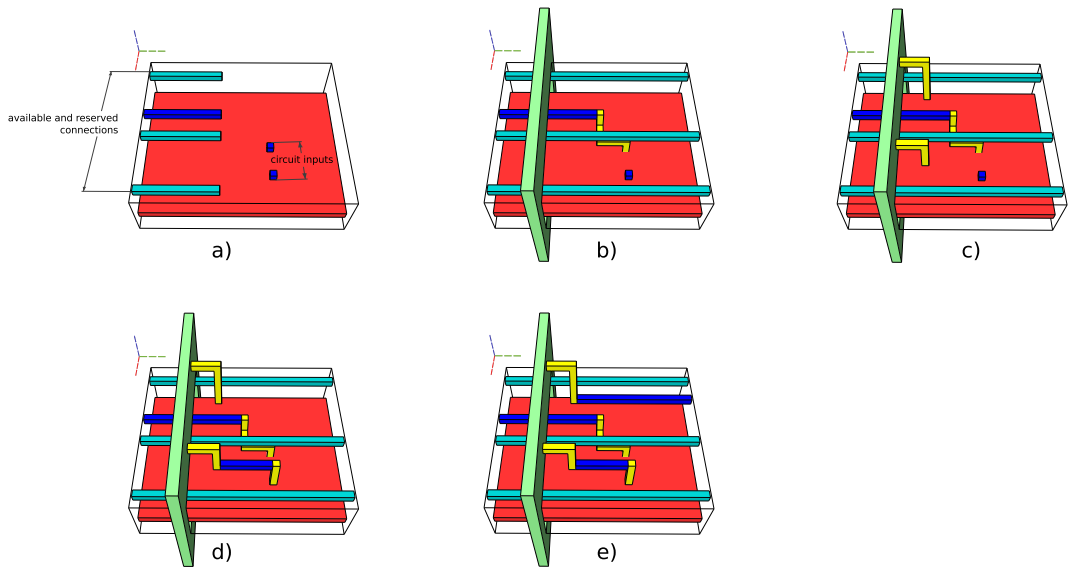

**Figure 8.** Connection pool example. The connection pool is parallel to the geometric description (red cuboid), and contains two types of connections. a) There are three connections of the same type (light blue) and a single connection of another type (dark blue). Circuit traversal has stopped, because two dark blue inputs are required. The manager is queried, and signals that only one high fidelity state is reserved; b) Scheduling is performed, the reserved state is assigned to one of the inputs, and the light blue connections are extended; c) Two boxes are successfully executed, and their outputs are connected to the pool (yellow). d) One of the new high fidelity states is assigned and connected to the second circuit input; e) The resulting extra dark blue connection is kept in the pool and extended.

There are two options if no distilled state is available: a) either delay initialisation and wait until boxes are scheduled and successfully executed, or b) interrupt synthesis with an error. Scheduling could delay the execution of circuit. because initialisations may have to wait for the execution of sufficient distillation boxes. Irrespective of the amount of available computing resources, there is a minimum time necessary for box execution. The exemplified scheduling does not delay circuit execution, but works in a proactive manner: place and execute sufficient, but not too many, boxes, so that there is a high probability that circuit inputs can use existing distilled states. This approach was used in Fig. 2. However, having too many distilled states available increases hardware resource requirements (see Discussion section).

Consequently, scheduling is performed if a certain *condition* is met, and the framework knows at any time the number of previously executed and successful distillation boxes. The scheduling condition can be related to the number of available successful boxes, to the time point a computation has reached, or any other meaningful criteria. In any synthesis configuration, the condition should be chosen to guarantee up to a high degree that each circuit initialisation can be performed. Fig. 2 uses a temporal condition.

Scheduling has to be resource efficient, and this aspect requires a trade-off analysis. It is envisioned, that quantum hardware will have a significantly higher cost compared to execution time. Thus, at least for the current generation of schedulers, the hardware footprint of the scheduled boxes has to be minimised, while time overhead is not limited. Fig. 2 uses the optimised ICM Toffoli circuit which has a longer time depth compared to the unoptimised one. However, hardware footprint is reduced by scheduling the boxes around a *thinner* geometric description.

**Connection manager**
The information about available successful distillation box outputs is managed by a dedicated component. The manager is queried to determine if scheduling would be necessary, and manages the available high fidelity states from the space-time volume. The manager is responsible for a so-called *connection pool*: in Fig. 2, a portion of the pool is the highest vertical layer of blue defects (abstracted as the blue cuboid from Fig. 5). The manager connects distillation boxes to the pool, extends available and unused connections to the latest time point of circuit traversal, and connects the circuit to the pool each time a high fidelity state is required. An example is offered in Fig. 8.

**Connection of distillation boxes to circuit initialisations**
ICM initialisations consume output states of successful distillation boxes. To this end, each such box needs to be connected to the geometry of the circuit by a pair of defects. In terms of the quantum circuit formalism this represents extending the output wire of the distillation sub circuit to the main circuit. In Fig. 3 the connection defects are yellow and connect the connection pool to the boxes (green in Fig. 4) and the circuit (red defects in Fig. 3).

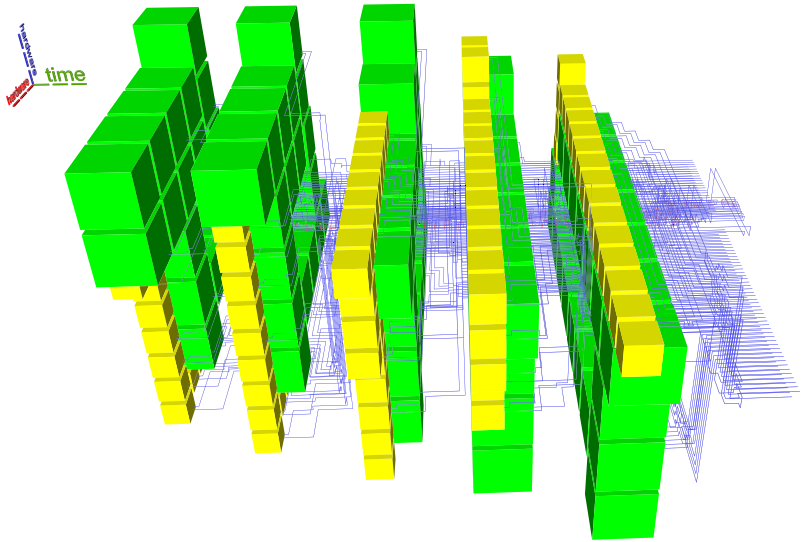

**Figure 9.** The number of connections in the pool is not limited, and increases after each scheduling round. The space-time volume of the entire circuit is greatly influenced by the width of the connection pool. Optimal synthesis has to predict and limit the maximum number of available connections.

The dynamic structure of the ICM geometric structure can be contextualised better, because the computation of connection defects is an intrinsic part of scheduling event handling. Connection dynamics is influenced by the results of box scheduling and execution: the exact location of the box output to use is known only after box execution. Therefore, connection defects are constructed differently in each circuit execution: the same ICM circuit input could be connected each time to a different box output. Dynamic path finding is a requirement for computing connection defects, and this offers another argument for the online synthesis of topological assemblies.

### Heuristic parameters

The online scheduling method is a heuristic controlled by a set of parameters (e.g. Fig. 5). The connection pool is placed above the circuit geometric description, at a parameterisable distance. The connection manager can be restricted to handle a maximum number of connections of each type. The time distance required for constructing the $\kappa_b$ segments, as well as the time distance allowed for connecting $\kappa_c$ is also parameterisable. Furthermore, both types of obstacles have a parameterisable minimum length. For some segments, this is automatically adapted to the number of elements in $\tau$.

Additionally, the currently implemented scheduler (spiral scheduler, see Methods section) is constructed around a *connection channel* surrounding both the connection pool and the circuit geometry. The channels are setup before synthesis and automatically increase with the number of the connections managed. For example, Fig. 9 illustrates the effect of proactively scheduling too many distillation boxes in each layer, and storing all successful states in the connection pool. The number of available connections constantly increases, because the circuit includes fewer high fidelity initialisations than the pool is manager is holding. The number of pool connections is not limited, and this leads to an increased hardware consumption (the width of the blue defect layer).

In general, there is not an optimal set of heuristic parameter values that can be used for the synthesis of an arbitrary quantum circuit. An optimal set of values would generate a minimal (at least efficient) bounding box of the entire synthesised computation. The software uses parameter values chosen after a sequence of synthesis iterations. The minimum values for which synthesis succeeds are chosen. Synthesis failures take place, because connection segments could not have been constructed, or not sufficient successful boxes were scheduled.

## Discussion

Previous state-of-the-art synthesis into topological assemblies consisted of two methods. The ASAP (as soon as possible) variant (e.g. Fig. 10 and Fig. 11) is an unoptimal scheduling method, where all the distillation boxes are placed before any element of the circuit geometric description is generated. The ALAP (as late as possible) variant (e.g. Fig. 12) places sufficient boxes along the circuit's geometric description time axis in order to guarantee that ICM initialisations can be executed.

| Step | Nr. A | Nr. Y | A Pool | Y Pool | Sched. Round |
|------|-------|-------|--------|--------|--------------|
| 1    | 2     | 1     | 6      | 6      | 1            |
| 2    | 0     | 1     | 4      | 5      | 0            |
| 3    | 0     | 1     | 4      | 4      | 0            |
| 4    | 1     | 0     | 4      | 3      | 0            |
| 5    | 0     | 1     | 3      | 3      | 0            |
| 6    | 0     | 0     | 8      | 8      | 1            |
| 7    | 1     | 0     | 8      | 8      | 0            |
| 8    | 0     | 1     | 7      | 8      | 0            |
| 9    | 1     | 0     | 7      | 7      | 0            |
| 10   | 0     | 0     | 9      | 9      | 1            |
| 11   | 0     | 2     | 9      | 9      | 0            |
| 12   | 1     | 0     | 9      | 7      | 0            |
| 13   | 1     | 0     | 8      | 7      | 0            |
| 14   | 0     | 0     | 8      | 10     | 1            |
| 15   | 0     | 1     | 8      | 10     | 0            |
| 16   | 0     | 1     | 8      | 9      | 0            |
| 17   | 0     | 1     | 8      | 8      | 0            |
| 18   | 0     | 1     | 8      | 7      | 0            |
| 19   | 0     | 1     | 8      | 6      | 0            |
| 20   | 0     | 1     | 10     | 7      | 1            |
| 21   | 0     | 1     | 10     | 6      | 0            |

**Table 1.** The number of $|A\rangle$ and $|Y\rangle$ states required after each circuit traversal step (Nr. A and Nr. Y), and the number of unused connections from the pool after each scheduling round (A Pool and Y Pool). The maximum number of connections is limited to ten (e.g steps 14, 20).

The new scheduling achieves a more compact placement compared to the ASAP and ALAP schedulers. To demonstrate this by example, the same optimised ICM Toffoli gate was synthesised using ASAP, ALAP and the new method. The equivalent volumes of the bounding boxes encompassing the topological assemblies was computed as the number of required plumbing pieces[15] to represent the assembly. The new method achieves a volume of 556920 plumbing pieces, which is about 30% less volume compared to the best state-of-the-art result (ALAP). The other obtained equivalent volumes are presented in figure captions.

The correctness of the synthesis result requires an analysis of the underlying ICM circuit. The optimised ICM version of the Toffoli gate implementation consists of nine wires, requires seven $|A\rangle$ and 14 $|Y\rangle$ high fidelity states. While traversing the circuit, the maximum number of $|A\rangle$ and $|Y\rangle$ states required simultaneously is two (see Table 1). Therefore, each distillation round includes a sufficient number of boxes (in this example, 14 of each type), such that accounting for their failure probability (in this example, 50%), the maximum number of successful boxes is guaranteed (two, in this example) for each traversal step. Table 1 illustrates how the number of unused connections from the connection pool (columns *A Pool* and *Y Pool*) fluctuates after each scheduling round (column *Sched* has value one) and after each circuit traversal step (columns *Nr. A* and *Nr. Y* have non-zero values).

Two additional views of the same synthesis result using the newly developed synthesis framework are presented in Fig. 13, and 14. The connection pool is placed above the ICM circuit in Fig. 13, and distillation boxes are spiralling around pool and circuit. In Fig. 14 the ICM geometric description is below the connection pool and cannot be easily identified. The analysis of these two figures suggests that future optimisation could consider methods to better control the number of distillation boxes in each round. For example, the volume of the assembly in Fig. 9 is 951048. This is almost as much as the volume obtained by the not optimal ASAP, and significantly more than the ALAP result. Thus, although Fig. 2 and Fig. 9 refer to the same ICM circuit, they have drastically different volumes. However, this shows that it is possible to further reduce the hardware footprint of a computation while not affecting the availability of high fidelity states in the connection pool.

## Methods

This section presents the algorithms used in each of the framework's components, and details the overall online synthesis work flow. The parameters influencing the synthesis results, and the supported diagnosis features are discussed.

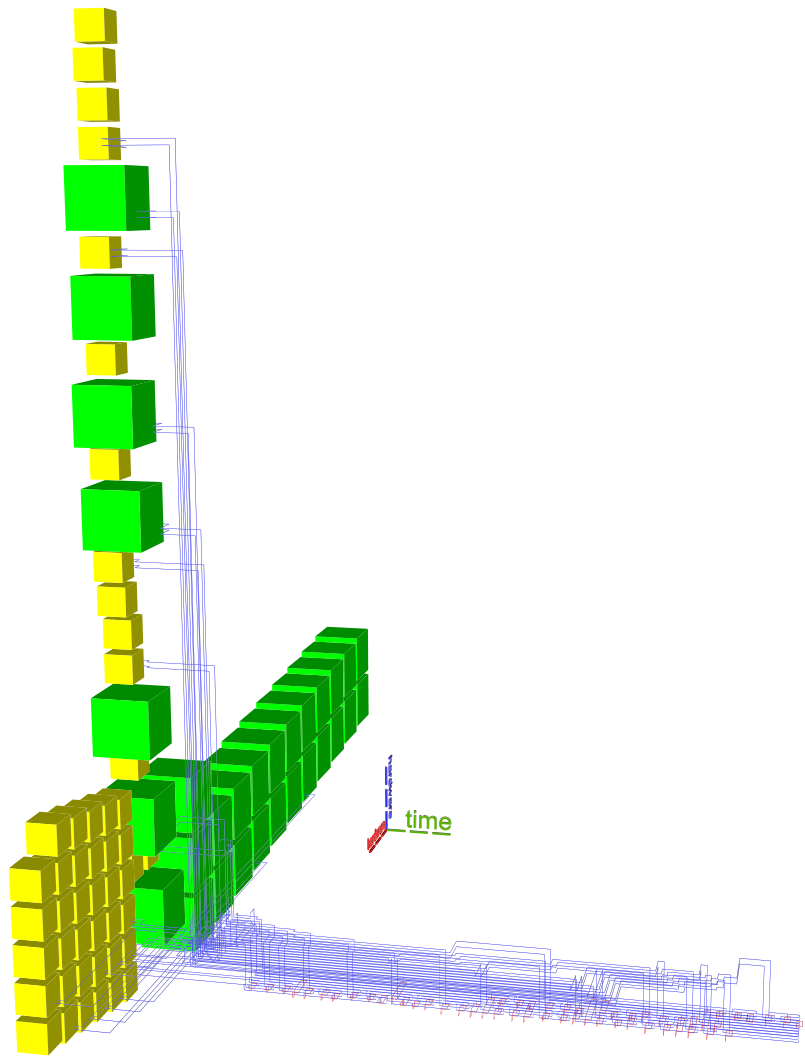

**Figure 10.** Previous state-of-the art example. ASAP scheduling using the optimised Toffoli circuit. The reduced number of wires generates a disproportionately high stack of the boxes. This assembly requires a total of 3336284 plumbing pieces.

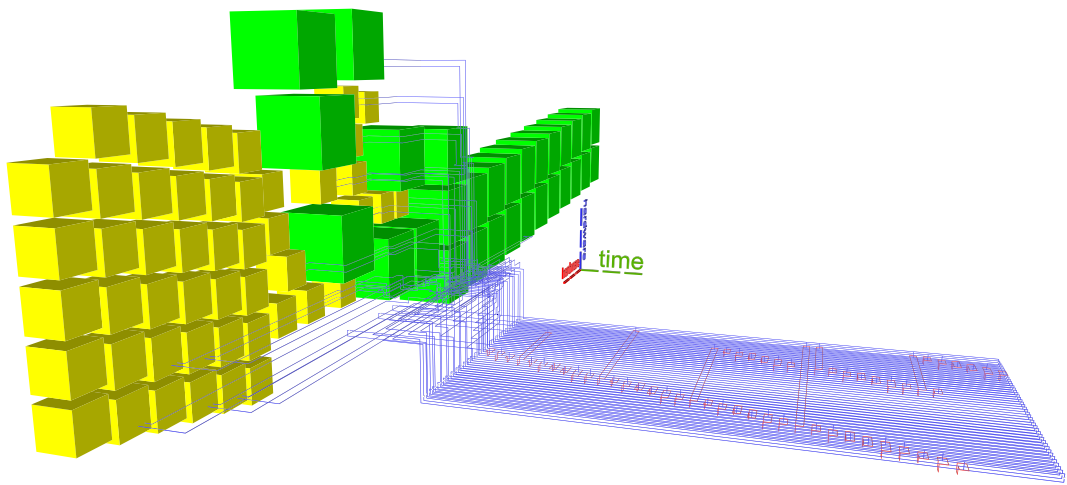

**Figure 11.** Previous state-of-the art example. ASAP scheduling using the unoptimised Toffoli circuit. This assembly occupies about one third of the one from Fig. 10 (1156896 plumbing pieces). Thus, ICM circuit optimisation does not always result in lower overall space-time volume if care is not taken where distillation boxes are placed.

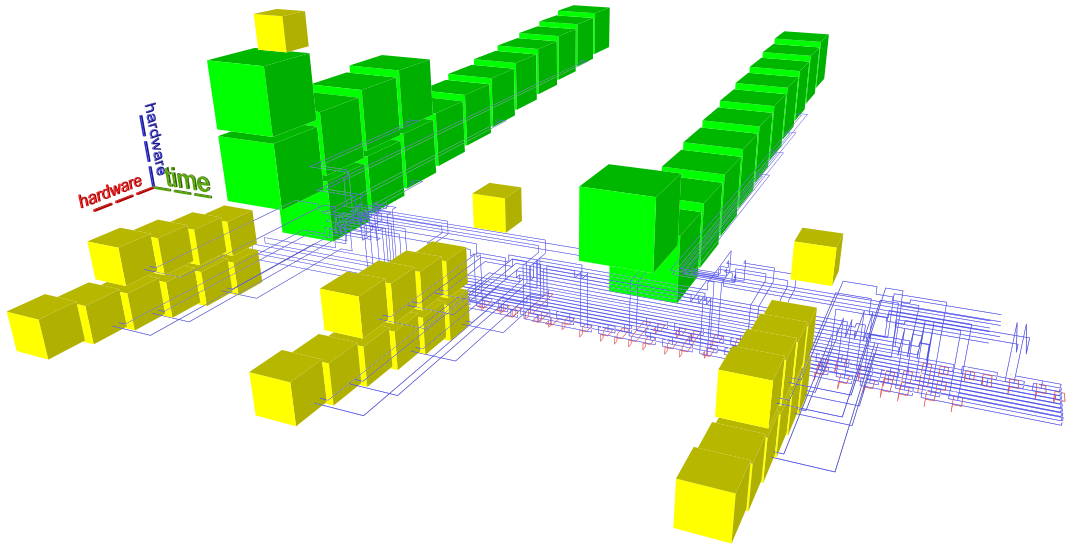

**Figure 12.** Previous state-of-the art example. ALAP scheduling using the optimised Toffoli circuit achieves a total of 798252 plumbing pieces.

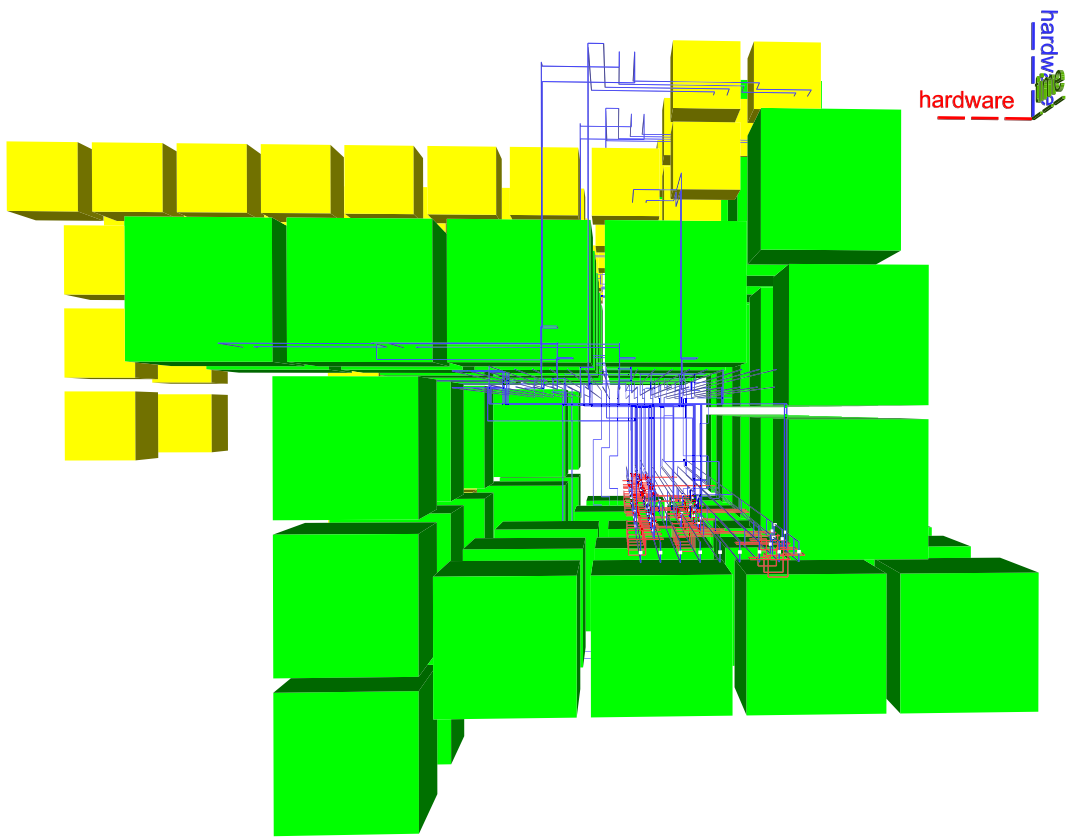

**Figure 13.** A different view of the synthesis result from Fig. 2. The time axis runs from back to the front. The connection pool is above the ICM circuit geometry, and boxes are scheduled counter-clockwise around them. The most visible layer of boxes is the last one scheduled during the circuit's execution.

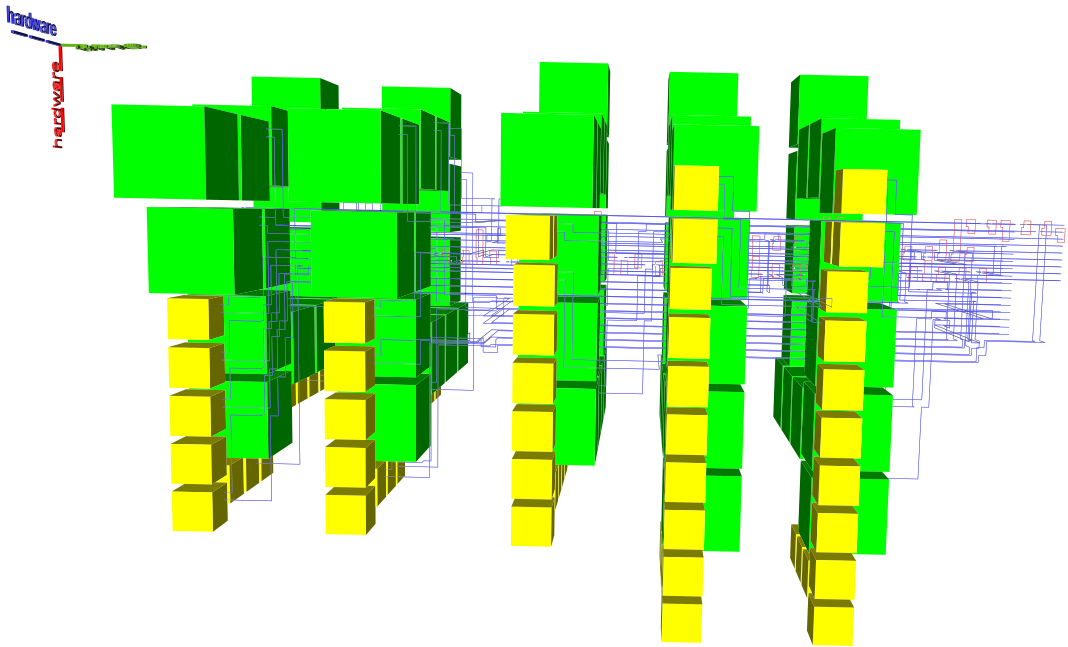

**Figure 14.** A different view of the synthesis result from Fig. 2. The time axis runs from left to the right. The connection pool is above the ICM circuit geometric description. The right-most layer of boxes is the one visible in Fig. 13. Each layer has the same number of boxes, but is differently layed out. This is because the width of the channel surrounding the connection pool increases with the constantly increasing number of connections managed (see Heuristic section). Therefore, from the first layer only five yellow boxes ($|Y\rangle$ distillations) are visible, whereas in the last two layers there are ten visible yellow boxes. These ten yellow boxes are the top-most boxes from Fig. 13.

### Connection manager operation

The connection manager is a framework component managing the geometric description of a connection pool. The connection pool is placed in a space-time region parallel to the one of the ICM circuit geometric description. In the current version the pool is placed above the circuit. The two regions (circuit and connection pool) are separated by a configurable distance.

The connection pool consists of multiple rails, which will be occupied by connections (pairs of defects). All rails are straight and parallel among themselves. They are also parallel to the circuit. After distillation box execution, each successful box is dynamically associated to a rail. The latter will be occupied by the connection originating from the box. Connections are kept in the pool (extended along the rail) until the circuit needs a specific initialisation state. At this point the connections are finalised: they are descended from the pool to the corresponding circuit pins.

A connection can be in only one of multiple states (see state transition diagram in Fig. 15). Each available distilled output is connected to an *available* connection (an unused one). For example, one can assume that there are six connections in the figure: four reserved, and two available. Available connections are simply represented by the empty space where these can be placed (e.g. Fig. 8a)).

The manager *reserves* the connection and assigns it the type of the particular distilled output. Available connections do not have a type, but reserved ones are of $|A\rangle$ or $|Y\rangle$ types. In the same figure, once the back yellow connection is constructed between box layer and pool, it has a certain type coded by dark blue.

The manager checks if there are sufficiently reserved, once a circuit input requires an $|A\rangle$ or $|Y\rangle$ distilled state. If that is the case, it *assigns* the connection(s) to the input(s) (Fig. 8b). If not, the manager can decide to trigger a scheduling event or an error (the green cuboid in Fig. 8b). Once the distilled output is connected by a pair of defects to the circuit geometry, the connection is marked as *tobeavailable*. This is done in order to guarantee that a pool rail is still occupied although the rails is marked available. The manager repeatedly checks each *tobeavailable* rail and resets the *available* ones. For example, the connection which was available in a) is connected to the circuit, thus theoretically freeing the rail. However, the back yellow connection cannot use the recently freed rail (Fig. 8c), because a connection segment is still occupying it.

### Spiral scheduling algorithm

Distillation box placement should deliver out of the box good enough results with respect to the occupied space-time volume. A solution is to place boxes in a spiral around the circuit's geometric description. This generates a high density region of boxes. Sequential scheduling events generate separate spirals, and the overall density of the space-time volume is dictated by the density of the connection segments. For illustrative purposes see Fig. 2.

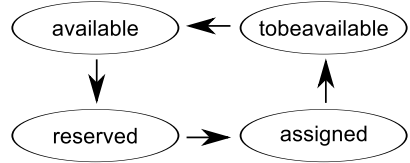

**Figure 15.** State transition diagram of a connection.

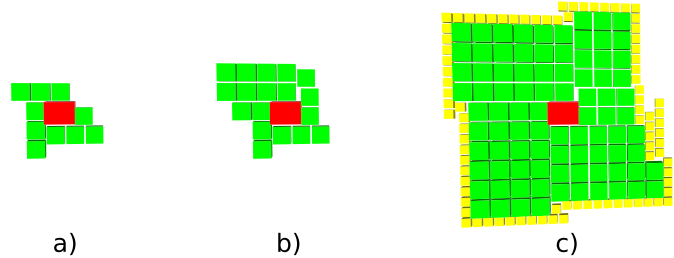

**Figure 16.** Spiral scheduling example. Green and yellow boxes are spiralled around the red box. Few boxes (e.g. a)) do not generate a balanced arrangement, i.e. the red box is not in the geometric centre. However, for large number of boxes the spiral pattern converges to a balanced arrangement.

Each spiral scheduling round consists of two phases: a) *calibration*; b) *individual box placement*. The coordinates of the first box to be scheduled are computed during calibration. The first distillation box is placed by assuming no previous knowledge except the requested time point. The box position is determined based on the time point (coordinate in the space-time volume) when scheduling occurs, and by iteratively performing *collision detection* with all other bounding boxes of distillations, connection pool and circuit geometry (see Section about heuristic parameters). The scheduler probes each potential placement against a framework central data structure (based on R-trees[22]). The data structure maintains references to all bounding boxes existing in the circuit's generated space-time volume. Individual boxes are placed at coordinates related to the previously placed box. A dense spiral layout is obtained by performing collision detection for each individual placement (e.g. Fig. 16).

## Computing connections

A connection is a pair of defects starting at the output of a successfully executed distillation box and ending at the geometric coordinates of the circuit logical qubit requiring an $|A\rangle$ or $|Y\rangle$ state initialisation. The ICM circuit, for which the geometric description is generated, is traversed in a sequence of steps triggering scheduling events. Handling each event generates a set of connection computation tasks, denoted $\tau$ including each connection $\kappa$ to be computed or updated. An A*-based[23] path finding algorithm is used.

Fig. 8 illustrates this discussion. Each connection consists of three segments: $\kappa_b$ (yellow, between box and connection pool), $\kappa_e$ (blue, along the connection pool rail), and $\kappa_c$ (yellow, from the pool to the circuit). Each segment is computed using a separate call to the path finding routine.

### *Connection model*

An arbitrary connection segment $\kappa_i^\pi \in \tau$ is defined as a tuple $(start_i, stop_i, obstacles_i, \pi)$, where start and stop are the coordinates of the segment endpoints. The set obstacles is used to better control the path finding routine, and $\pi$ is an integer value representing a priority.

The path finding routine uses a geometric distance heuristic (similar to Manhattan distance). The heuristic influences the direction of the segments forming the connection. However, it cannot be guaranteed that a previously computed connection will not block the path of a future connection. Obstacles are used for this reason and they represent *no go areas* in the space-time volume. There are *guide* and *occupy* obstacles: guides are applicable to all computed segments/connections, and occupies only for the one currently computed. Therefore, a guide $\gamma$ of a particular connection $\kappa_i$ is valid for all connections $\kappa \in \tau$. On the contrary, the occupy $\omega$ of connection $\kappa_i$ will affect only $\kappa_i$ and none of $\kappa_j \in \tau$, for $i \neq j$.

Connection priority $\pi$ indicates the order in which connections are computed. For example, the connection $\kappa^1$ (having $\pi = 1$) has lower priority and is computed after $\kappa^9$ (having $\pi = 9$). Obstacles inherit the connection priority, but obstacle priorities have an inverse meaning: lower numbers are more important. For example, for some connections with priorities $3\ldots1$, the highest connection priority is 3, and the lowest is 1. Therefore, connections are computed in the order $3,2,1$. However, obstacle priorities are inverted: if the obstacles of 3 and 1 intersect in the space-time volume, the obstacle 1 will override obstacle 3 at the intersection. This is performed in order to increase the guarantee of connections not blocking each other: the number of available coordinates (unused by previous paths) from the space-time volume decreases with each computed connection. Therefore, the last $\kappa \in \tau$ (having the lowest priority) needs to be protected using obstacles. For example, in Fig. 17 the yellow $\omega$-obstacle of the black connection has a higher priority than the orange $\omega$-obstacle of the grey connection.

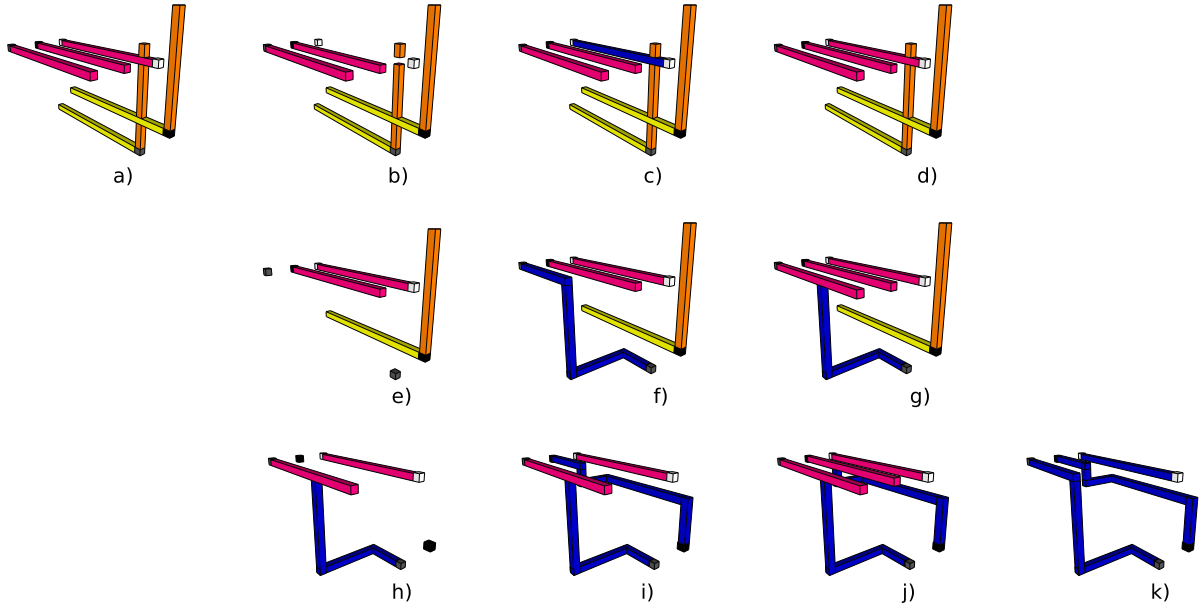

**Figure 17.** Computing three connection segments using obstacles. a) The start and end coordinates are marked by white, grey and black cubes. The connection priorities are: white (255), grey (128) and black(0). Therefore, the white connection segment is computed first, and the black one last. Magenta bars indicate $\gamma$-obstacles, and yellow and orange bars $\omega$-obstacles. Blue bars represent the defects forming the connection spanned between the start and stop coordinates. b) all obstacles of the white connection are disabled; c) the defects of the connection are computed; d) the $\gamma$-obstacle is re-enabled; e-g) computation of the grey connection; h-j) computation of the black connection; k) final result after all $\gamma$-obstacles are disabled.

At the same time, any $\gamma$-obstacle has a higher priority than an $\omega$-obstacle. For example, in the same figure, the $\gamma$-obstacle of the white connection has a higher priority than the orange $\omega$-obstacle of the grey connection.

A segment computation consists of two steps: a) determining the values of each $\kappa$ tuple (start and stop coordinates, necessary obstacles and its priority); b) path finding between start and stop considering the obstacles. For computing an arbitrary $\kappa_i$, the obstacles are manipulated as follows: 1) before computation every $\gamma$ and $\omega$-obstacle from the corresponding set of $\kappa_i$ is disabled; 2) after computation only $\gamma$-obstacles are re-enabled. The latter operation is necessary for protecting the connection pool rails, so that each connection is guaranteed a linear range of space-time volume coordinates. As a result, $\gamma$-obstacles are, for the moment, mostly used for protecting connection pool rails.

It should be noted that connection priorities and obstacle configurations need to be carefully chosen. The example in Fig. 17 is not guaranteed to work if priorities are kept but the type of the orange $\omega$-obstacles would be changed to $\gamma$: the grey connection could still block the black one.

*Connection segment order*

Connection segments are determined in the order: $\kappa_c$, $\kappa_e$, $\kappa_b$. This ordering is chosen because two scheduler types are possible: 1) influencing the circuit geometry; 2) not influencing circuit geometry. The first type could arise when hardware resources are very scarce, and parallel execution of distillations is not supported: circuit execution would need to be halted until sufficient distilled states are available, the circuit's geometry being dynamically influenced. The latter scheduler type is preferred in this work. An example of how schedulers are influencing the geometry, and thus the coordinates of the places where distilled states are to be connected, is presented in Fig. 18.

The suitability of the same connection segment determination order for both scheduler types is shown in the following.

Let *In* be the set of all circuit inputs of the circuit being synthesised. The circuit's operations impose a time ordering on all the inputs. The following input types exist: the ones already connected to distillation boxes ($In_a$), the ones having a distillation box output assigned but not connected ($In_c$), inputs for which scheduling was triggered ($In_b$) and not been assigned any distillation box output, and future inputs ($In_f$).

After each circuit traversal step, the space-time volume coordinates of the inputs from $In_a$ and $In_c$ are known. This is not the case for $In_b$ and $In_f$ inputs, where the time coordinates are only assumed. With each circuit traversal the size of $In_a$ increases by $|In_c|$, but it is not guaranteed that the one of $In_f$ decreases by $|In_b|$. Circuit traversal is concerned only with $In_f$, but $|In_f|$ and the time ordering between its inputs is influenced by the scheduler type (again, see Fig. 18).

A scheduler influencing the geometry could force a time delay $\Delta$ of the $In_b$ inputs. In contrast, for schedulers not influencing the geometry the delay is $\Delta = 0$. After scheduling, irrespective of scheduler type, the inputs in $In_b$ are delayed by $\Delta$, and a new time ordering of $In_f$ is generated. The time order of inputs in $In_a$ and $In_c$ is not affected. Inputs triggering scheduling events act like a time-border between *known* and *assumed time coordinates*. Therefore, one can connect an input to the connection pool, only when it precedes such a scheduling triggering border input.

As a conclusion, input connection segments $\kappa_c$ can be determined easily for inputs from $In_c$, while $In_b$ are not considered for connections until they are not returned from $In_f$ in future steps. Connection extension segments in the connection pool ($\kappa_e$) can be determined after each traversal step. Changing the state of assigned connections into tobeavailable is necessary.

Connection segments cannot be constructed to $In_b$ inputs, because no corresponding connections were reserved in the first place, due to the missing boxes. Segments $\kappa_b$ (segments connecting successfully executed boxes to the pool) can be determined after: a) pool connections to $In_c$, and b) $\kappa_e$ were determined. An example of how segments are connected to the pool and to the circuit is presented in Fig. 8.

| iter # | connected | assigned | scheduling | future |
|--------|-----------|----------|------------|--------|
| 1      | ①         | ② ③      | ④          | ⑤      |
| 2a     | ① ② ③     |          | ⑤          | ④      |
| 2b     | ① ② ③     | ④        | ⑤          |        |

**Figure 18.** Inputs ordering as a result of using different schedulers. During the first step the circuit consists of five inputs, of which number 1 was previously connected to a distilled state, numbers 2 and 3 need to be connected, number 4 triggered a scheduling round, and the fifth is not processed. If scheduling would introduce a delay, input 4 could be moved into the future beyond input 5 (situation 2a). The result is that input 5 would be scheduled before input 4 is connected. If scheduling does not delay inputs, the next traversal step would have to connect input 4 and schedule input 5.

### Synthesis work flow

Circuit synthesis is a repeating sequence of steps: circuit traversal, box scheduling, geometric description, connection computation. The following algorithm sketch illustrates the overall synthesis method.

The implemented synthesis triggers a next scheduling event right after the current scheduling round has finished. This is performed in order to minimise the risk of forcing the circuit geometrical description to be delayed. This approach introduces an additional source of probability in the synthesis, but it is possible to minimise failure by pre-analysing the temporal distribution of inputs along the circuit's execution.

### Synthesis diagnosis

Diagnosis of synthesis failures requires a software component for logging all relevant operations performed. The current implementation uses a simplistic component that synchronises the state of the active obstacles and of the used *plumbing pieces*[15] (units of a defect) from the space-time volume. The synthesis state right before failure can be visualised. The plumbing piece state is kept in a text journal allowing, in case of failures, to diagnose/debug the fault.

Online synthesis will require a more potent diagnosis method, including means to online verify the synthesised/optimised quantum computation. However, the current diagnosis component was sufficient for the capabilities of the implemented synthesis.

## Conclusion

This work introduced a novel framework for the synthesis of topologically error-corrected quantum circuits which includes

- an event driven and structured synthesis method;
- an online distillation box scheduler;

```
 1: circuit ← Read ICM circuit
 2: geom ← Geometric description constructor
 3: scheduler ← Scheduling method for distillation box placement
 4: manager ← Connection manager
 5: Optimise circuit
 6: Traverse circ until first unscheduled input
 7: while circuit traversal not finished do
 8:     if manager does not have enough distilled states then
 9:         scheduler schedules missing distillation boxes
10:         Simulate successful distillation boxes
11:         manager reserves connections for successful boxes
12:     end if
13:     time ← the minimum time coordinate of an unscheduled input
14:     Generate circ geometry until time
15:     Assign a connection to each input whose coordinates were computed
16:     Determine the $\kappa_c$ connection segments: from manager to circuit inputs
17:     Mark connections involved in $\kappa_c$ as tobeavailable
18:     Determine the $\kappa_e$ connection segments: extensions along pool rails
19:     Determine the $\kappa_b$ connection segments: from boxes to manager.
20:     Add necessary obstacles to all $\kappa_b$, $\kappa_e$, $\kappa_c$
21:     Compute segments in the order $\kappa_c$, $\kappa_b$, $\kappa_e$
22:     Check which tobeavailable connections are indeed free at time, and mark them as available
23:     Traverse circuit until first unscheduled input
24: end while
```

- a path finding method for computing defect connections between distillation boxes and circuit inputs;

- a set of parameters for controlling synthesis;

- a preliminary diagnosis component;

The current method was demonstrated using the ICM Toffoli gate implementation. Compare to the best previous state-of-the-art result, the new method achieved a space-time volume reduction of approximately 30%. This was possible by placing the distillation boxes in a systematic and online manner. There is potential to further optimise these results. First, heuristic parameter values influence directly the achieved bounding box volumes. Methods to minimise values (e.g. the time distance parameter) could prove beneficial. Second, synthesis could be based on probabilistic models of the circuit in order to dynamically evaluate and predict resource (number of distillation boxes) consumption. Future work will also research methods to improve the performance of the framework.

## Acknowledgement
A.P. acknowledges support from the Linz Institute of Technology grant LITD13361001.

## Author contributions statement
A.P. and A.G.F. conceived the idea. A.P. was responsible for algorithm development and simulations. A.G.F. and R.W. were responsible for results verification. All authors were responsible for drafting of the manuscript.

## Competing financial interests
The authors declare no competing financial interests.

## Data availability
No datasets were generated or analysed during the current study. Source code is available online at the link indicated in the abstract.